# How the learning environment predicts male and female students' motivational beliefs in algebra-based introductory physics courses

Sonja Cwik, Kyle Whitcomb, and Chandralekha Singh
*Department of Physics and Astronomy, University of Pittsburgh, Pittsburgh, PA 15260*

Societal stereotypes and biases pertaining to who belongs in physics and who can excel in physics can impact motivational beliefs, e.g., of women and racial and ethnic minority students in physics courses. This study investigates how the learning environment predicts male and female students' motivational beliefs including physics self-efficacy, interest, and identity at the end of year long (spanning two-semester) algebra-based introductory physics courses. These were courses at a large university in the US taken primarily by biological science majors many of whom are interested in health professions. Although women are not underrepresented in these physics courses, societal stereotypes and biases internalized by female students over their lifetime can still impact their motivational beliefs about physics. Our findings show gender gap in motivational beliefs favoring men. These findings can be useful to provide support and create an equitable and inclusive learning environment to help all students excel in these courses.

## I. INTRODUCTION, FRAMEWORK AND GOAL

Women are underrepresented in many science, technology, engineering, and math (STEM) careers and prior research has focused on understanding issues pertaining to their representation and participation in these fields [1-4]. Studies have shown that motivational beliefs such as students' identity and self-efficacy in a particular field are important for the students' continuation in those STEM majors and careers [5-9]. Many studies have focused on calculus-based physics courses, in which women tend to be severely underrepresented and it has been found that women on average have lower self-efficacy than men [10-12]. Few prior studies have focused on physics self-efficacy and identity in algebra-based introductory physics courses for bioscience majors in which women are the majority.

However, pervasive stereotypes and biases about who belongs in physics and who can succeed in physics could impact women, even in these algebra-based physics courses. One common stereotype is that genius and brilliance are important factors to succeed in physics [13]. Genius is often associated with boys [14] and girls from a young age tend to shy away from fields associated with innate brilliance or genius [15]. Studies have found that by the age of six, girls are less likely than boys to believe they are "really really smart" and less likely to choose activities that are made for "brilliant people" [15]. These stereotypes continue to impact women throughout K-12 education and in college [16].

Moreover, students' identities in STEM disciplines plays an important role in their in-class participation and choices of majors and careers [17-19]. Prior studies show that it is more difficult for women to form a physics identity than men [20,21]. Additionally, our model stems from prior work that shows that a student's physics identity is influenced by their physics self-efficacy, interest and perceived recognition [21-24]. Self-efficacy in a discipline refers to students' belief in their ability to accomplish tasks or solve problems. It has been shown to impact students' engagement, learning, and persistence in science courses [23,25-27]. For example, when tackling difficult problems, students with high self-efficacy tend to view the problems as challenges that can be overcome whereas those with low self-efficacy tend to view them as personal threats to be avoided [27]. Similarly, interest in a discipline can affect students' perseverance and achievement [28-30], so changing the curriculum to conform to female students' interests led to improved understanding of all students [31].

Our study focuses on the premise that due to societal stereotypes and biases, male and female students' physics motivational beliefs, including self-efficacy, interest, and identity, may be different even in a course in which women are not underrepresented. Investigating their perceptions of the learning environment can provide guidelines for how to make learning environments equitable and inclusive to ensure that all students develop high motivational beliefs and learn physics effectively. In particular, at the end of the year long algebra-based introductory physics courses (physics 1 and 2), we investigated male and female students' perceptions of the learning environment, which include their perceptions of being recognized as a physics person by others (including friends, family and their instructors/teaching assistants or TAs), their interactions with their peers and whether they felt they belonged in the physics class. These factors were included since students' sense of belonging in physics have been shown to correlate with students' self-efficacy [32,33] and retention. Similarly, students' interaction with peers has been shown to enhance engagement and perceived recognition has been shown to play an important role in students' identity [34]. The perception of learning environment includes experiences students have in the classroom as well as interactions outside of the classroom like office hours of the instructor or TA and students studying or doing homework together. We investigated physics self-efficacy, interest and identity at the end of physics 2 to answer the following research questions:

**RQ1** Are there gender differences in students' physics self-efficacy, interest and identity at the end of the course and do they change from physics 1 to physics 2?

**RQ2** How does the perception of the learning environment predict motivational factors at the end of the course?

## II. METHODOLOGY

We administered a validated written survey at a large public research university in the U.S to students at the end of the semester (post) in traditionally taught introductory algebra-based physics 1 and physics 2 over the course of two years. We used data from 544 matched students who completed the survey on paper scantrons (47% response rate) in the last two weeks of recitation class who were enrolled in physics 1 if taken in the fall and physics 2 if taken in the spring. The classes are typically taken by students in their junior or senior year majoring primarily in biological sciences with approximately 50%-70% of students expressing a desire to pursue future careers in health professions. The university provided demographic information such as age, gender, and ethnic/racial information using an honest broker process by which the research team received the information without knowledge of the identities of the participants. From the university data, the participants were 37% male and 63% female. We recognize that gender is not a binary construct. However, the data provided by the university only included binary options (less than 1% of the students did not provide this information and thus were not included in the analysis).

The survey instruments were constructed from items validated by others [35-37] and re-validated by us in our own context using individual interviews [10], exploratory and confirmatory factor analyses (CFA) [38], Chronbach alpha [39] and Pearson correlation between different constructs [38]. The survey items asked about different aspects of students' motivational beliefs at the time it was administered and the perception of their learning environment. These motivational constructs include students' physics identity (1 item), self-efficacy (4 items), interest (4 items), perceived recognition (3 items), sense of belonging (5 items), and interaction with their peers (4 items). The CFA was conducted to establish a measurement model for the constructs and used in SEM. In the CFA, the model fit indices were good and all of the factor loadings were above 0.50, which indicate good loadings [38]. Pearson's correlations $r$ vary in the strength of their correlations for different pairs of constructs; however, the largest correlation was 0.89 and therefore all of the factors could be considered separate constructs [38].

The *physics identity* items evaluated whether the students see themselves as a physics person [9]. The *physics self-efficacy* questions measured students confidence in their ability to answer and understand physics problems [40]. The *interest in physics* items measured student enthusiasm and curiosity to learn physics [40]. The *sense of belonging* items focused on whether students felt like they belonged in the class [33,36]. The *perceived recognition* items measured the extent the student thought that other people see them as a physics person [9]. Lastly, the *peer interaction* items measured whether students thought that working with their peers was beneficial to them [41,42].

The survey items in this investigation were on a Likert scale of 1 (low endorsement) to 4 (high endorsement) except for sense of belonging items which were designed on a scale of 1 to 5 to keep them consistent with the original survey [43]. The rating scales for the specific items varied in order to provide a more valid measure of intensity of response (e.g., Strongly Disagree, Disagree, Agree, Strongly Agree). A lower score (1) was indicative of a negative endorsement of the survey construct while a higher score (4 or 5) was related to a positive endorsement. Some of the items were reverse coded (e.g., I feel like an outsider in this class).

We first analyzed descriptive statistics and compared female and male students' mean scores on various constructs for statistical significance using *t*-tests and computed the effect sizes using Cohen's *d* [38]. For predictive relationships between different motivational constructs, we used Structural Equation Modeling (SEM) as a statistical tool, which involves a measurement part (e.g., CFA) and a path analysis. The SEM used R (lavaan package) with a maximum likelihood estimation method [44]. SEM is an extension of multiple regression. It conducts several multiple regressions simultaneously between variables in one estimation model and can have multiple outome variables. It allows calculation of overall goodness of fit and allows for all estimates to be standardized simultaneously so that there can be direct comparison between different structural components along with calculations of factor loadings for all factors (or constructs or latent variables). In the year long course, we used matched data from physics 1 and 2 in that all students have responded to the survey questions at the end of both physics 1 and 2. We controlled for self-efficacy and interest at the end of (post) physics 1 to predict how the perception of learning environment effects post self-efficacy, interest and identity in physics 2. We report model fits for SEM using the Comparative Fit Index (CFI), Tucker-Lewis Index (TLI), Root Mean Square Error of Approximation (RMSEA), and Standardized Root Mean Square Residuals (SRMS). Commonly used thresholds for goodness of fit are these: CFI and TLI > 0.90, SRMR and RMSEA < 0.08 [45].

The model estimates were first performed using gender moderation analysis to check whether any of the relations between variables show differences across gender by using "lavaan" to conduct multi-group SEM. Initially we tested different levels of measurement invariance in the model. In each step, we fixed different elements of the model to equality across gender and compared the results to the previous step using the likelihood ratio test. Since we did not find statistically significant moderation by gender, we tested the theoretical model in a gender mediation analysis, using gender as a variable directly predicting all latent variables to examine the resulting structural paths between constructs.

### III. RESULTS AND DISCUSSION

To answer RQ1, Table 1 shows that the mean values of all constructs for men and women are statistically significantly different in favor of men. This pattern is similar to that found in calculus-based physics courses [24,46,47] at the end of physics 2, despite the fact that women are the majority in the algebra-based physics courses (63%). One hypothesis for the gender gap in these constructs in Table 1 (indicated by their lower scores) is that women may be affected by previous experiences, stereotypes and biases about who belongs in physics and who can excel in physics, which may have accumulated over their life time.

However, we do find one significant difference in our findings here for the algebra-based physics courses, in which women are not underrepresented. In particular, in Table 1, for both post self-efficacy (gray) and post interest (blue) constructs that have scores from both physics 1 and physics 2, we find no statistically significant differences in the scores from physics 1 to physics 2 (i.e., these post scores are stable from physics 1 to physics 2). However, in the calculus-based courses, in which women are severely underrepresented, the gender gap in these constructs becomes larger from physics 1 to physics 2 [10,11]. Thus, in algebra-based courses, while the inequity signified by the mean value of the gender gap in these constructs is not getting worse from physics 1 to physics 2, it is not improving either. Since students' self-efficacy, interest and identity are important for their continuation in STEM fields, it is critical to make the physics learning environment equitable and inclusive to help bridge the gap.

With regards to the predictive model and to answer RQ2 (see Fig 1), we used SEM and investigated the relationships between different constructs. We initially tested gender moderation between different constructs using multi-group SEM (between female and male students) and investigated whether any of the relationships between the variables differed across gender. There were no group differences at the level of weak and strong measurement invariance including no difference at the level of regression coefficients. Therefore, we proceeded to gender mediation analysis to understand how gender mediates student outcomes through the perception of the learning enviornment (physics self-efficacy, interest, and identity) at the end of the year long introductory physics sequence, controlling for students' post self-efficacy and interest at the end of physics 1 and the learning environment (see Fig. 1).

In the SEM model (Fig 1), in addition to gender, we included students' perceived recognition, peer interaction, and belonging as perceptions of the learning environment to mediate student outcomes (post physics 2 self-efficacy, interest and identity). All paths were considered and nonsignificant lines were cut ($p > 0.05$) except for the path from interest to identity in line with our theoretical framework. The model fit indices indicate a good fit to the data: CFI = 0.928, TLI = 0.918, RMSEA = 0.058, SRMR = 0.049. Fig 1 shows that there is no direct effects from gender to any of the post student outcomes in physics 2, i.e., post self-efficacy, interest, and identity. Moreover, we find that gender only had direct connections to self-efficacy (the strength of the relationship given by the standardized regression coefficient $\beta = 0.28$) and interest ($\beta = 0.36$) at the end of physics 1. This is in line with what we find for the mean values in Table 1: no differences between the interest and self-efficacy scores of women and men from physics 1 to physics 2. Therefore, much of the difference in male and female students' motivational constructs are carried over from physics 1 to physics 2 and could potentially be due to the societal stereotypes and personal experiences pertaining to who belongs in physics.

Fig.1 shows that interest in physics 2 has the largest direct connection from interest in physics 1 ($\beta = 0.78$) with smaller effects from peer interaction ($\beta = 0.16$) and perceived recognition ($\beta = 0.12$). Although, in our model, interest in physics 2 is mainly correlated with interest in physics 1, it does not mean that interest can not be changed over the time in the class. Our model shows that learning environment could predict interest through the small contribution from peer interaction and perceived recognition. One possible way to improve students' interest in physics is to provide more problems in class that relate to students' interests and career paths and help students discern the relevance of physics.

**Table 1** Mean values of the constructs as well as statistical significance (p-values) and effect sizes (Cohen's d) by gender. p-values are indicated with no superscript for p <0.001 and with "a" for p = 0.002. There were 345 women and 199 men in this sample. In the table, S.E. is self-efficacy, P.R. is perceived recognition, Peer Int. is peer interaction, Bel. is belonging, and Phys Idt. is physics identity.

| Predictors and Outcome | Mean | | Cohen's d |
|---|---|---|---|
| | Male | Female | |
| S.E. in Phys 1 (1-4) | 2.98 | 2.73 | 0.49 |
| Interest in Phys 1 (1-4) | 2.81 | 2.38 | 0.71 |
| P.R. in Phys 2 (1-4) | 2.23 | 1.98 | 0.38 |
| Peer Int. in Phys 2 (1-4) | 2.96 | 2.79 | 0.28[a] |
| Bel. in Phys 2 (1-5) | 3.76 | 3.47 | 0.34 |
| S.E. in Phys 2 (1-4) | 2.93 | 2.72 | 0.41 |
| Interest in Phys 2 (1-4) | 2.77 | 2.30 | 0.76 |
| Phys Idt. in Phys 2 (1-4) | 2.18 | 1.85 | 0.45 |

Fig. 1 also shows that self-efficacy in physics 2 has direct effects from self-efficacy in physics 1 ($\beta = 0.41$), belonging ($\beta = 0.35$), peer interaction ($\beta = 0.21$), and a small effect from perceived recognition ($\beta = 0.16$). Self-efficacy is important for students' persistence in the class and future career choices. Since the learning environment can influence students' self-efficacy, it is important for an instructor to try and improve students' perceived recognition and sense of belonging as well as improve the peer interaction by creating an equitable and inclusive classroom in which students are excited to interact with peers without the fear of being judged if they are wrong. Instructors can also influence students' peer interaction by providing time for the students to work together in class and making sure all voices are heard equally when discussing problems as a whole group.

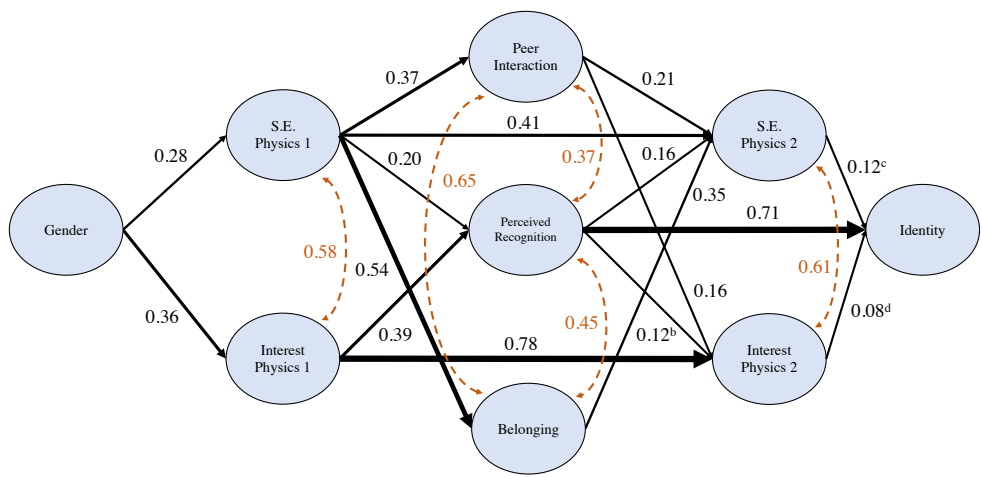

**Fig 1** Result of the SEM with gender mediation. Students' post self-efficacy, interest and identity outcomes in physics 2 are predicted by the perceptions of learning environment (perceived recognition, peer interaction and belonging) holding gender as well as post self-efficacy and interest in physics 1 as controls. The line thickness qualitatively denotes the relative magnitude of the standardized regression coefficients β shown. The dashed lines are covariances between factors. All *p*-values for β are indicated by no superscript for *p* < 0.001, "b" for *p* = 0.003, "c" for *p* = 0.004, and "d" for *p* = 0.056.

Furthermore, Fig 1 shows that self-efficacy and perceived recognition influence physics identity in physics 2 directly, with perceived recognition having the largest direct effect ($\beta = 0.71$). Additionally, Fig. 1 shows that there is a small direct effect from interest to physics identity. Also, prior research in calculus-based courses in which women are underrepresented shows that compared to men, women have more negative experiences pertaining to perceived recognition from the instructors and TAs. In Table 1, we see that even in the algebra-based physics courses in which women are not underrepresented, both women and men have a mean recognition below the positive lower threshold (score of 3) and women are lower then men. Furthermore, in Table 1, women also have lower scores on identity. Therefore, it is important for instructors and TAs to meaningfully recognize their students even in the algebra-based physics courses.

## IV. IMPLICATIONS AND FUTURE DIRECTIONS

Our model shows that factors comprising student perception of learning environment are important for explaining student outcomes of physics self-efficacy, interest, and identity at the end of physics 2. Instructors have the ability to influence these factors (students' perceived recognition, sense of belonging, and peer interaction), and can empower all their students by making their classes equitable and inclusive. These factors also influence each other, so if an instructor improves students' peer interaction by allowing students to work in groups during class in an inclusive manner, it could also increase students' sense of belonging. If instructors can provide support for one of the factors comprising learning environment they can most readily control (eg., peer interaction) and make their classroom more equitable, they are likely to improve student outcomes in the process.

Ultimately we want an equitable and inclusive learning environment in all physics courses. If the physics 2 classes discussed here were like that, what would the SEM model in Fig. 1 look like? Envisioning an ideal (utopian) situation can help to understand what the model should look like as the classroom becomes increasingly more inclusive and equitable. The physics 2 instructor has no control over the motivational factors or other variables before the students enter their classroom. Therefore there would be no differences in the mean values for each of these variables and there would be no changes to the regression lines for anything before the physics 2 variables. However, in an ideal situation, at the end of the physics 2 class, all students would have the top score in every motivational factor, no matter what their scores were in the physics 1 factors. This would mean that every student would have a strong self-efficacy in physics, would feel that they belonged in the classroom, etc. Therefore, all regression lines to and from the learning environment factors would be clipped (regression coefficient zero or not statistically significant). This is because students who had poor scores in physics 1 would be able to excel at the end of physics 2 and there would be no difference in the motivational factors to distinguish these students at the end. With regards to the gender variable, there would be an extra regression line going to all the learning environment and outcome variables that favors women to make up for their different scores in the controlled variables. While this situation is not realistic, it sheds light on how our current situation in physics 2 is not reflecting an equitable and inclusive learning environment.

There are a variety of ways that TAs/instructors can improve student interactions and the learning environment in the courses. One way to improve the learning environment is through classroom interventions [48-50]. Brief social-psychological classroom interventions, e.g., mindset and sense of belonging interventions, have been shown to increase women's sense of belonging and self-efficacy in the classroom, as well as boost their confidence and interest in physics [4,48-50]. At the same time, these interventions can help students develop positive feelings of being recognized by their peers, TAs, aAnd instructors. We have implemented a 25 minute belonging intervention in the calculus-based physics classes that has been shown to eliminate the gender gap in physics performance [49]. The intervention could be adapted for the algebra-based physics courses as well.

In summary, more can be done in the physics classrooms to mitigate the stereotypes and past experiences women have had over their life time that perpetuate through to the end of the students' introductory physics courses. It is important for instructors and TAs to provide a learning environment that emphasizes recognizing their students, allowing for positive peer interactions and providing a space where all students can feel that they belong in physics. From our analysis, these factors play a central role in predicting students' self-efficacy, interest, and identity in physics. It is important to note that student perception of the learning environment is not shaped only by what happens in the classroom. Student interactions with each other while they do homework, students' experiences in an instructor's or TA's office hours, interactions between students and the instructor over email and other circumstances all contribute to the students' learning environment. All of those interactions can affect students' identity, self-efficacy, and interest in physics.

## V. ACKNOWLEDGEMENTS

This work was supported by grant NSF DUE-152457.